\documentstyle[epsfig,12pt]{aipproc}

\def \beq{\begin{equation}}
\def \eeq{\end{equation}}
\def \efi{Enrico Fermi Institute Report No. EFI}

\begin{document}
\title{Theoretical Issues in the Tevatron Era
\footnote{Invited talk presented at Symposium in Celebration of the
Fixed-Target Program with the Tevatron, Fermilab, June 2, 2000, to be published
in Comments on Modern Physics.  \efi~2000-24, hep-ph/0007194.}}
\author{Jonathan L. Rosner}
\address{Enrico Fermi Institute and Department of Physics \\
University of Chicago, Chicago, IL 60637 USA}
\maketitle
\begin{abstract}
The Fermilab Tevatron's operation for fixed-target physics from its start
in 1983 until the end of fixed-target running in 2000 was marked by
extraordinary productivity and variety.  Some of the changing
theoretical issues associated with this program are reviewed.
\end{abstract}

\section{Introduction}

The Fermilab Tevatron was constructed with several aims in mind.  (1) It
would permit the storage of counter-rotating beams of protons and
antiprotons at beam energies of up to 1 TeV, allowing for collisions at
center-of-mass energies approaching 2 TeV.  (2) It would demonstrate the
first large-scale use of superconducting technology.  (3) It would allow
the Fermilab energy for fixed-target programs, which had typically been
400 GeV, to be doubled to 800 GeV, and would permit a saving of
electrical power.  In this last context the Tevatron project was known
as the ``Doubler/Saver.''  The present article is devoted to
theoretical issues accompanying the Fermilab fixed-target Tevatron
program.

Table 1 outlines some topics and how they evolved during the 17 years of
Tevatron fixed-target operation, from its inception in 1983 until the
end of fixed-target running in 2000.  In many cases, theoretical questions
have been totally transformed by the advent of precision measurements at
the Tevatron and elsewhere.  In others, new discoveries have raised as
many questions as they set out to answer.

In 1983 the $W$ and $Z$ had just been observed, while we had only a
hazy idea of where to look for the top quark, and the tau neutrino was
anticipated but not yet seen.  Now, with precise measurements of $W$
and $Z$ masses and couplings, and a top quark mass known to a better
fractional accuracy than that of any other quark, we can begin to
anticipate the Higgs boson mass.  For a single Higgs in the Standard
Model, the best value comes out tantalizingly close to present lower
limits \cite{Quigg}.  The discovery of the tau neutrino has been reported
\cite{DONUT}.

In 1983, the $b$ quark lifetime had just been shown to lie in the range
of 1 ps \cite{blife}, with the $b \to c$ coupling hence surprisingly
small and the $b \to u$ coupling even smaller \cite{WP}.
The asymmetry in the beta-decay $\Sigma^- \to n
e^- \bar \nu_e$ \cite{oldSig} disagreed with the prediction \cite{Cab}
of the Cabibbo theory of semileptonic hyperon decays.  Charmed particle
lifetimes were starting to be mapped out \cite{oldcharm} but theoretical
understanding of them was primitive \cite{GLR,GLAM}.  CP violation in the
neutral kaon system, nearly twenty years after its discovery \cite{CCFT},
still could be parametrized by a superweak $\Delta S = 2$ interaction
\cite{SW} mixing $K^0$ and $\bar K^0$. The proposal \cite{KM} that phases
in weak couplings of quarks were responsible for this effect was still
many years away from being confirmed.  Today, we are close to mapping out
both magnitudes and phases of weak quark couplings
\cite{CKMrevs}; the beta-decays of
$\Sigma^-$ and other hyperons confirm the Cabibbo theory \cite{E715};
the hierarchy of charmed particle lifetimes is at least qualitatively
understood \cite{Cumalat}; and direct CP violation (as predicted by the
Kobayashi-Maskawa theory) has been observed in the form of a difference
between the CP-violating ratios $\Gamma(K_L \to \pi^0 \pi^0)/\Gamma(K_S
\to \pi^0 \pi^0)$ and $\Gamma(K_L \to \pi^+ \pi^-)/\Gamma(K_S \to \pi^+
\pi^-)$ \cite{E832,NA48}.

\renewcommand{\arraystretch}{1.2}
\begin{table}
\caption{Evolution of theoretical topics during the Tevatron era.}
\begin{center}
\begin{tabular}{l l l}
Topic & \qquad 1983 & \qquad 2000 \\ \hline
Electroweak    & $W,~Z$ seen & Precision measurements, $m_t$ \\
~~symmetry and & No top      & ~~constrain Higgs mass \\
~~neutrinos    & No $\nu_\tau$ & Discovery \\ \hline
Weak quark     & $\tau(b) \sim 1$ ps! & Precise CKM elements \\
~~couplings    & $\Sigma^- \to n e^- \bar \nu_e$? & Cabibbo confirmed \\
               & $\tau$(charm)? & Hierarchy understood \\
               & Direct CP violation? & $\epsilon'/\epsilon \ne 0$ \\
 \hline
QCD and        & EHLQ $q(x)$ & Quark and gluon structure \\
~~hadron       & Gluons?     & ~~functions with error bars \\
~~structure    & SSC plans   & LHC construction under way \\
               & $Q \bar Q$ production? & Some progress \\
               & Hyperon polarizations? & Some progress \\ \hline
The unexpected & Magnetic monopoles     & Supersymmetry \\
               & Neutral heavy leptons  & ~~(but is it really \\
               & Toponium               & ~~unexpected?) \\
               & Compositeness & \\
               & 1.8 MeV $e^+ e^-$ bump & \\ \hline
\end{tabular}
\end{center}
\end{table}

Many problems in QCD and hadron structure were addressed in the
Tevatron era.  In 1983 the proton structure functions of Eichten,
Hinchliffe, Lane, and Quigg \cite{EHLQ} helped anticipate physics at
supercolliders.  The mechanism for quarkonium production was unclear,
but a combination of direct QCD effects and electromagnetic cascades
from higher levels seemed possible.  Hyperons were found to be produced
with polarizations which depended on transverse and longitudinal
momenta and hyperon species.  We are now the beneficiaries of greatly
improved knowledge about both quark and gluon distributions in hadrons,
for example thanks to precise neutrino deep inelastic scattering studies
at Fermilab and elswhere \cite{CCFR},
and are looking forward to reliable error estimates for these functions.
Although the Superconducting Supercollider (SSC) did not survive
funding cuts, the Large Hadron Collider (LHC) is on track toward
operation during the middle of this decade. New experiments
have solved some mysteries of quarkonium production and hyperon
polarization, but uncovered others.

During the Tevatron era the attitude of many physicists toward the
``unexpected'' may have become less flexible. In 1983 the
possibilities for new physics seemed richer and less universally
agreed upon than they do today.  Searches were under way for magnetic
monopoles, neutral heavy leptons, toponium, quark and lepton
compositeness, and even an elusive bump at 1.8 MeV in the $e^+ e^-$
spectrum.  Today, although some of these searches have even been pursued
recently, many physicists seem apologetic if they are not looking for
the odds-on favorite among most theorists, supersymmetry.  The field thus
seems somewhat more monolithic than it was in 1983.  Part of the great
advantage of the Tevatron was the opportunity it provided for a rich
variety of experiments on a scale that could be managed by collaborations
with modest resources but original ideas.  One hopes to see future
possibilities for this variety of approaches.

Roughly 45 fixed-target experiments were performed using the Tevatron
during the period 1983--2000. These are summarized in Table 2;
more details may be found in Ref.~\cite{FT}. We shall touch upon some
aspects of this program from the theoretical standpoint.

We describe progress on electroweak symmetry and neutrinos in Section
2, and on weak quark couplings in Section 3.  Section 4 is devoted to
charmed particle lifetimes, while Section 5 treats charm mixing and CP
violation.  Section 6 deals with other results on heavy quarks.  Two
topics in hadron structure, the magnetic moments of baryons and the
polarization of hyperons, are reviewed in Sections 7 and 8.  Some
possibilities for unexpected physics are mentioned in Section 9, while
Section 10 concludes.

\section{Electroweak Symmetry and Neutrinos}

\subsection{Precise measurements}

The ratio $R_\nu$ of the rate of neutral-current to charged-current
interactions of neutrinos was one of the first sources of information
about the weak mixing angle $\theta_W$, defined in lowest
electroweak order as $\sin^2 \theta_W \equiv 1 - (M_W/M_Z)^2$.
This relation has continued to define $\theta_W$ in the
presence of electroweak radiative corrections, while a slightly
different quantity $\theta_{\rm eff}$ is measured through 
precise studies of $Z$ couplings at LEP and SLC.  Small differences
between the two arise as a result of loops involving, for example,
the top quark and the Higgs boson.

\begin{table}
\caption{Fixed-target experiments at the Fermilab Tevatron, 1983--2000.}
\begin{center}
\begin{tabular}{l l c}
Topic        & Subtopic            & Fermilab expt.\  no.\ \\ \hline
Kaons        & CP violation        & 621, 731, 773, 832 \\
             & Rare decays         & 799 (+ hyperons)   \\ \hline
Hyperons     & $\Sigma^- \to n e^- \bar \nu_e$ & 715 \\
             & $\mu(\Omega^-)$     & 756, 800 \\
             & Radiative decays    & 761 \\
             & Charmed baryons     & 781 \\
             & CP violation        & 871 \\ \hline
Neutrinos    & Counter             & 733, 744, 770, 815, 872 \\
             & Bubble chamber      & 632, 745 \\ \hline
Hadron       & Muon scattering     & 665, 782 \\
\quad structure & Direct photons   & 706 \\
             & Hadron jets         & 557/672, 609, 683 \\
             & Polarized scattering & 581/704 \\
             & Structure functions & 866 \\ \hline
Pair         & Dimuons             & 605, 615, 772 \\
\quad spectrometers & Dihadrons    & 605, 711, 789 \\ \hline
Heavy        & Charm(onium)        & 400, 653, 687, 690, 705, \\
\quad quark  &                     & \quad 743, 769, 791, 831 \\
\quad production & Beauty          & 690, 771, 789 \\ \hline
Particle     & 1.8 MeV$/c^2$       & 774 \\
\quad search & \quad $e^+e^-$ bump & \\ \hline
\end{tabular}
\end{center}
\end{table}

From a simplistic viewpoint, which is a slight distortion of
the actual situation, the neutral-current cross section $\sigma_{NC}
(\nu N)$ involves $Z$ exchange, and the $Z$ boson's mass is well
measured, while the charged-current cross section $\sigma_{CC}
(\nu N)$ involves exchange of the $W$, whose mass is less well
measured.  Thus, measurement of $R_\nu$ serves mainly to constrain
$M_W$, whatever the mass of the top quark or Higgs boson.  A similar
conclusion applies to a combination of neutral-current and
charged-current cross sections known as the Paschos-Wolfenstein
\cite{PW} ratio,
\beq
R_{PW} \equiv \frac{\sigma_{NC}(\nu N) - \sigma_{NC}(\bar \nu N)}
{\sigma_{CC}(\nu N) - \sigma_{CC}(\bar \nu N)}~~~.
\eeq
The NuTeV Collaboration \cite{NuTeV} has used this ratio to determine
$M_W = 80.26 \pm 0.11$ GeV (for nominal $m_t,~M_H)$.
Given the known value of the top quark mass
\cite{mt}, $m_t = 174.3 \pm 5.1$ GeV, this value of the $W$ mass can
be combined with other direct measurements to constrain the Higgs
boson mass $M_H$ to lie below about 200 GeV.

\subsection{The tau neutrino}

The tau neutrino is the one fermion in the Standard
Model that remains to be observed directly.  Confirmation
would not only cement our confidence in the many indirect
measurements that require its existence, but also would help us
learn how to see $\nu_\tau$'s in experiments which seek to study
their appearance in oscillations.

A number of years ago a beam dump experiment was proposed \cite{bd}
in order to produce and study $\nu_\tau$'s.  Cost and schedule
constraints prevented its implementation.  The main source of
$\nu_\tau$'s was expected to be the decay $D_s^+ \to \tau^+
\nu_\tau$.  Since then, the $D_s$ decay constant has been measured
by several groups including one at Fermilab \cite{Kodama},
with the most precise value \cite{CLEODs} implying a favorable
branching ratio for this process of about 6\%.
As of June 2000 the DONUT experiment (Fermilab E-872) had several
$\nu_\tau$ candidates, for which results now have been published \cite{DONUT}.

\subsection{Neutrino oscillations}

The evidence that neutrinos have mass and undergo oscillations from
one species to another includes several results:

\begin{itemize}

\item Neutrinos from the Sun appear at the Earth with a probability
ranging from about 30 to 60\% of that expected, depending on their
energies \cite{solex}.  The interpretation of this effect in terms
of neutrino oscillations \cite{solth} allows for several ranges of mass
differences and mixing angles.

\item Muon neutrinos produced in the atmosphere appear to oscillate
into another species, most likely $\nu_\tau$, with near-maximal
mixing $\sin^2 2 \theta \simeq 1$ and $\Delta m^2 \simeq 3 \times
10^{-3}$ eV$^2$ \cite{atm}.

\item One experiment \cite{LSND} has presented evidence for the
oscillation $\bar \nu_\mu \to \bar \nu_e$, with an allowed region in
$\sin^2 2 \theta$ -- $\Delta m^2$ resembling a sinking canoe.

\end{itemize}

The confirmation and elaboration of the second and third of these
results is an important part of Fermilab's future fixed-target program
\cite{MINOS,BooNE}.

\section{Weak Quark Couplings}

The pattern of charge-changing couplings of the weak quarks is as much
a fundamental mystery as the masses of the quarks, and probably springs
from the same physics.  It is expressed in terms of the unitary
Cabibbo-Kobayashi-Maskawa (CKM) \cite{Cab,KM} matrix $V$, which has four
real parameters for three families of quarks.  These may be taken to be
\cite{WP} (1) $\lambda = \sin \theta_C \simeq 0.22$, where $\theta_C$ is
the Cabibbo angle \cite{Cab,GL}; (2) $A = |V_{cb}|/\lambda^2 \simeq
0.8$; (3) $\rho = {\rm Re}(V_{ub}/A \lambda^3) \simeq 0.0$ -- 0.3;
(4) $\eta = - {\rm Im}(V_{ub}/A \lambda^3) \simeq 0.3$ -- 0.5.  The
existence of a nonzero complex phase in some elements of $V$ was a
crucial ingredient in Kobayashi and Maskawa's
explanation of CP violation in the neutral kaon system.  It required
the existence of the third quark family, both of whose members were
discovered at Fermilab \cite{Ups,top}.  With only two families of
quarks, a single parameter $\lambda$ would have sufficed to describe
the mixing \cite{Cab,BG}.

Fermilab has played a major role in measuring elements of the CKM
matrix, either directly or via loop effects.  Neutrino and
charm experiments confirm the expectations of a unitary $V$
by showing that $|V_{cs}| \simeq 1$,
$|V_{cd}| \simeq |V_{us}| \simeq 0.22$ \cite{Cumalat,CCFR}.
Collider results on $B$ meson and top quark decays have improved our
information on $V_{cb}$ and confirm (within wide limits)
that $V_{tb} \simeq 1$ as expected from unitarity.  Hyperon
($\Sigma^-$, $\Lambda$, $\Xi^0, \ldots$) beta decays \cite{E715,hypbeta}
confirm that $|V_{us}| \simeq 0.22$ and provide insights into
the nature of SU(3) violations \cite{JM} in these processes.  Studies
of CP-violating $K^0$--$\bar K^0$ mixing \cite{E832} and collider
experiments on $B^0$--$\bar B^0$ mixing give information on the phase
and magnitude of $V_{td}$, while future collider experiments on
$B_s$--$\bar B_s$ mixing will constrain the ratio $|V_{ts}/V_{td}|$
and hence $|V_{td}|$, given our expectation that $V_{ts} \simeq -
V_{cb}$.

A more extensive discussion of the constraints on $V$ may be found in
Refs.~\cite{CKMrevs} and \cite{SJ}.  Here we mention only a few key
points.

\subsection{Direct CP violation in neutral kaon decays to two pions}

As noted in the Introduction, the definitive observation at
Fermilab of direct CP violation in neutral kaon decays \cite{E832}
has qualitatively validated the Kobayashi-Maskawa theory \cite{KM},
which was previously favored over the superweak \cite{SW} picture
just on the basis of the magnitudes of CKM matrix elements \cite{SJ}.  My own
average (June 2000) for the parameter Re($\epsilon'/\epsilon$) describing
this effect, based on experiments at Fermilab and CERN
\cite{E832,NA48,E731,NA31} is $(19.2 \pm 4.6) \times 10^{-4}$, where
I have included a scale factor to account for the poor agreement
among these very challenging measurements.  More data are expected
from both Fermilab and CERN.  The present world average is somewhat
above the favored range of theoretical predictions \cite{Buras}, but
uncertainties in hadronic matrix elements can probably account for
any discrepancy \cite{K99}.

\subsection{$K^+ \to \pi^+ \nu \bar \nu$}

The decay $K^+ \to \pi^+ \nu \bar \nu$ is sensitive mainly to the
top quark's contribution to a loop diagram, with a small correction
for charm, and so constrains the combination $|1.4 - \rho - i \eta|$.
One predicts \cite{BuK} a branching ratio
\beq
{\cal B}(K^+ \to \pi^+ \nu \bar \nu) \simeq 10^{-10} \left|
\frac{1.4 - \rho - i \eta}{1.4} \right|^2~~~,
\eeq
so that for $0 \le \rho \le 0.3$ one expects a branching ratio ${\cal
B} \simeq (0.8 \pm 0.2) \times 10^{-10}$, with additional errors
associated with the charmed quark mass and the Wolfenstein parameter
$A$.  A measurement of this branching ratio to 10\% could provide a
significant constraint on the parameter $\rho$ or could exhibit
interesting deviations from the Standard Model prediction.  At present
one event has been recorded by Brookhaven Experiment E787 \cite{E787},
corresponding to ${\cal B} = (1.5^{+3.4}_{-1.2}) \times 10^{-10}$.
More data are expected, both from further analysis of E787 and from
an approved follow-up experiment (E949) at Brookhaven \cite{E949}.
A Fermilab proposal \cite{CKM} also seeks to study this process.

\subsection{$K_L \to \pi^0 \ell^+ \ell^-$}

The decay $K_L \to \pi^0 \ell^+ \ell^-$ has important CP-violating
contributions, both direct (proportional to $\eta$) and indirect
(resulting from the admixture $\epsilon K_1$ of the CP-even state
in $K_L$).  Each contribution separately would give rise to a
branching ratio ${\cal B}(K_L \to \pi^0 e^+ e^-)$ of several parts in
$10^{12}$.  Background from the decay $K_L \to \gamma \ell^+ \ell^-$
in which a final lepton radiates an extra photon \cite{HG} may
limit the search for this process.  Moreover, the CP-conserving
process $K_L \to \pi^0 \gamma \gamma \to \pi^0 \ell^+ \ell^-$ also
probably plays a role at a significant level.  Present 90\% c.l.
upper limits \cite{pll} are ${\cal B}(K_L \to \pi^0 [e^+ e^-,~
\mu^+ \mu^-]) = [5.1,3.8] \times 10^{-10}$, a factor of about 100
above interesting levels unless the indirect contribution is far
greater than most estimates \cite{BuK,Ko}.

\subsection{$K_L \to \pi^0 \nu \bar \nu$}

The decay $K_L \to \pi^0 \nu \bar \nu$ is purely CP-violating and
provides a clean probe of $\eta$.  The predicted branching ratio,
proportional to $A^4 \eta^2$, is expected to be about $3 \times
10^{-11}$ \cite{BuK}.  The best upper limit on the branching ratio
utilizes the Dalitz decay of the $\pi^0$ and is $5.9 \times 10^{-7}$
\cite{pll}.  Proposals exist to improve these limits at Brookhaven
\cite{K0pio} and Fermilab \cite{KAMI}.

\subsection{Other rare kaon decays}

A recent study of the decay $K_L \to \mu^+ \mu^- \gamma$ \cite{BQ}
may help us to learn more about the CKM matrix (particularly the
parameter $\rho$) by pinning down long-distance effects due to the
two-photon contribution in $K_L \to \gamma \gamma \to \mu^+ \mu^-$.
Although it is not relevant to CKM physics, one should also mention
the observation, both at Fermilab \cite{KTeVa} and at CERN
\cite{NA48a}, of a CP- and/or T-violating asymmetry in the decay
$K_L \to \pi^+ \pi^- e^+ e^-$, in accord with theoretical predictions
\cite{SeS} based on the observed CP-violating mixing in the neutral kaon
system.

\section{Charm Lifetimes}

The lifetimes of charmed particles exhibit an interesting interplay
of short-distance and long-distance effects, intermediate between the
case of kaons, where long-distance effects dominate, and particles
containing $b$ quarks, where short-distance physics can explain most
(but not all) of the pattern.  Several effects must be taken into
account.

\begin{enumerate}

\item The rate for the decay of a free charmed quark, $c \to s + (u
\bar d, e \nu, \mu \nu)$ is given in terms of the muon decay rate
$\Gamma_\mu = 4.55 \times 10^5$ s$^{-1}$ by
\beq
\Gamma(c \to s + \ldots) \simeq 5 \left( \frac{m_c}{m_\mu} \right)^5
\Gamma_\mu \Phi~~~,
\eeq
where the phase space correction $\Phi$ is approximately (0.45,0.97)
for $m_c = 1.5$ GeV and $m_s = (0.5,0.1)$ GeV, leading to a predicted
lifetime $\tau(c) = (1.7,0.8)$ ps.  This is already in the ballpark
of the longest charmed particle lifetime, that of the $D^+$ (see
below).

\item A modest QCD enhancement of the subprocess $c \to s u \bar d$
is expected in the channel in which $s$ and $u$ form a color
antitriplet \cite{GLAM}.

\item Final states with two or more identical quarks can be subject to
either destructive or constructive ``Pauli interference''
\cite{Pauli}.

\item Long-distance (e.g., resonant) effects can enhance ``non-exotic''
channels, as they appear to do in the dominance of $\Delta I = 1/2$
weak nonleptonic decays of kaons and hyperons \cite{GLR}.  Thus, for
example, the lifetime of $K_S$, 0.089 ns, is much shorter than that
(12 ns) of the $K^+$.  The $K_S$ can decay to $\pi \pi$ in an $I = 0$ 
channel, for which the $\pi \pi$ interaction is strong (if not exactly
resonant).  By contrast, $K^+ \to \pi^+ \pi^0$ must be purely $I = 2$,
and there are no known resonances with $I = 2$.

\end{enumerate}

The Tevatron has been a major player in establishing the interesting
hierarchy of charmed particle lifetimes displayed in Table 3.  Evidence
for all of the above mechanisms seems to be present.  The key to these
studies has been the isolation of charmed particles in the presence of
formidable backgrounds by detecting their decays, only fractions of a
millimeter from their production, using silicon vertex detectors.

\begin{table}
\caption{Charmed particle lifetimes and effects contributing to them.}
\begin{center}
\begin{tabular}{l c c}
Particle & Lifetime (ps) \cite{PDG} & Comments \\ \hline
$D^+$    & $1.051 \pm 0.013$ & Exotic channel; Pauli int.\ lowers 
$\Gamma$ \\
$D^0$    & $0.4126 \pm 0.0028$ & Rate QCD-enhanced \\
$D_s^+$ & $0.496^{+0.010}_{-0.009}$ & Rate QCD-enhanced (suppr. by binding?) \\
$\Lambda_c^+$ & $0.206 \pm 0.012$ & Subprocess $cd \to su$ effective \\
$\Xi_c^0$ & $0.098^{+0.023}_{-0.015}$ & Subprocess $cd \to su$
effective; \\
          &                  & Pauli int.\ raises $\Gamma$ \\
$\Xi_c^+$ & $0.33^{+0.06}_{-0.04}$ & No subprocess $cd \to su$; \\
          &                  & Pauli int.\ raises $\Gamma$ \\
$\Omega_c^0$ & $0.064 \pm 0.020$ & Pauli int.\ raises $\Gamma$ \\
\hline
\end{tabular}
\end{center}
\end{table}

\section{Charm Mixing and CP Violation}

\subsection{$D^0$--$\bar D^0$ mixing and lifetime difference}

Both at Fermilab and elsewhere, there are new and potentially exciting
results on the $D^0$--$\bar D^0$ system.  The CLEO Collaboration
\cite{CLEOmix} has studied the time-dependence of ``wrong-sign''
decays $D^0 \to K^+ \pi^-$, thereby learning a combination of parameters
describing the mass difference $\Delta m$ and width difference $\Delta
\Gamma$ between the CP-even and CP-odd combinations of $D^0$ and $\bar
D^0$.  If one defines $\Gamma$ as the average width of these two states,
$x \equiv \Delta m/\Gamma$, $y \equiv \Delta \Gamma/\Gamma$, and
$\delta$ to be a relative final-state phase between $D^0 \to K^+ \pi^-$
and $\bar D^0 \to K^+ \pi^-$, the CLEO result entails
\beq
-5.8\% < y' \equiv y \cos \delta - x \sin \delta < 1\%~~~,
\eeq
hinting (though not with sufficient significance) at a negative value
of $y'$.

More recently, the FOCUS Collaboration (Fermilab E831) \cite{FOCUS}
has directly compared the lifetime of $D^0$ in the $K^- \pi^+$ mode,
which is half CP-even and half CP-odd, with that in the $K^+ K^-$ mode,
which is purely CP-even, finding
\beq
y = (3.42 \pm 1.39 \pm 0.74)\%~~~.
\eeq
Although the deviation from zero is not yet statistically compelling,
the central value is far larger than theoretical predictions
\cite{ChTh} and, if confirmed, could be a hint of new physics.

\subsection{CP violation}

CP-violating asymmetries in charmed meson decays are expected
to be small in the Standard Model.  Two factors contribute to the
expected smallness of penguin $c \to u$ amplitudes and $D^0$--$\bar
D^0$ mixing.  First, the internal $d$ and $s$ quark contributions
nearly cancel one another, as entailed in the original
Glashow-Iliopoulos-Maiani (GIM) mechanism \cite{GIM}.  Second, the
largest internal quark mass in the loop diagram for $c \to u$ is
$m_b$, which is much smaller than that ($m_t$) in the loop diagrams
for $s \to d$ or $b \to (d,s)$ transitions.  If we define
\beq
{\cal A}_{\rm CP} \equiv \frac{\Gamma(D) - \Gamma(\bar D)}
{\Gamma(D) + \Gamma(\bar D)}~~~,
\eeq
the FOCUS Collaboration \cite{FOCUSCP} finds
$$
{\cal A}(D^+ \to K^- K^+ \pi^+) = -0.006 \pm 0.011 \pm 0.005~~~,
$$
$$
{\cal A}(D^0 \to K^- K^+) = -0.001 \pm 0.022 \pm 0.015~~~,
$$
\beq
{\cal A}(D^0 \to \pi^+ \pi^-) = 0.048 \pm 0.039 \pm 0.025~~~.
\eeq
All these values are consistent with zero at the several percent level
and represent an improvement over previous bounds.

\section{Heavy Quark Results}

\subsection{$\Upsilon$ production}

The E605 Collaboration has studied the relative production in hadronic
reactions of $\Upsilon(1S)$, $\Upsilon(2S)$, and $\Upsilon(3S)$
\cite{E605}.  The relatively large height of the 3S peak leaves at
least some role for an electromagnetic cascade from the 3P states, which
are expected to lie below $B \bar B$ threshold \cite{Kwong}.

\subsection{$J/\psi$ and $\psi'$ production}

The production cross section of the $J/\psi$ in hadronic interactions
\cite{Jpsiprod} is too large to be accounted for purely by electromagnetic
cascades from the $\chi_c$ states \cite{GLR,EQ} or by direct production of the
color-singlet $c \bar c$ state.  A gluonic component of the wave
function, involving a color-octet $c \bar c$ state, seems to be
required \cite{Braaten}.  This should not be so surprising in view of
the fact that roughly half the nucleon's momentum is carried by gluons.
Difficulties still exist in explaining the production of the 2S $c \bar c$
state, the $\psi'$.  (For a recent optimistic discussion see Ref.\
\cite{Kniehl}.)  Is this because the 2S state has light quarks in its wave
function as a result of proximity to $D \bar D$ threshold?  Some problems must
remain for the next generation of particle physicists!

\subsection{$b$ quark production}

Many experiments at the Tevatron (e.g., \cite{E771,E789}) searched
for the production of $b$ quarks by 800 GeV protons on a fixed target.
The cross section is relevant for CP-violation studies now under way
at similar energies at HERA-b.  Using $J/\psi$'s identified as $B$
decay products by their displaced vertices, for example, the E789
Collaboration \cite{E789a} has measured $\sigma(pN \to b \bar b + X)
= 5.7 \pm 1.5
\pm 1.3$ nb/nucleon.  This value, while small, is larger than the
$e^+ e^- \to b \bar b$ cross section at the $\Upsilon(4S)$.

\section{Baryon Magnetic Moments}

Several experiments at Fermilab have added to our knowledge about
hadron structure through the measurement of baryon magnetic moments
\cite{magmom}.  The na\"{\i}ve quark model predicts that for a baryon
composed of two quarks $q_1$ of one kind and one $q_2$ of another,
the magnetic moment will be \cite{GR}
\beq
\mu(q_1 q_1 q_2) = \frac{4}{3} \mu(q_1) - \frac{1}{3} \mu(q_2)~~~.
\eeq
This prediction assumes that all quarks in the baryon are in a relative
S-wave.  A simple argument based on Fermi statistics and color then
demands that the two identical quarks $q_1$ be in a state of spin 1,
and the coefficients in the above equation are related to the
Clebsch-Gordan coefficients for coupling this system with the spin
of $q_2$ to obtain total spin of 1/2.

If the ground-state baryon contains configuration-mixed states
\cite{BS}, one can still obtain some predictions \cite{BR}.  The
proton and neutron mix differently with higher states than the
$\Sigma$ and $\Xi$ since decuplets of SU(3) contain states which
can mix with the latter but not with the former.  Thus, one writes
$$
\mu(q_1 q_1 q_2) = (A - B)\mu(q_1) + B \mu(q_2)~~~(p,n)~~~,
$$
\beq
\mu(q_1 q_1 q_2) = (A'-B')\mu(q_1) + B' \mu(q_2)~~~(\Sigma,\Xi)~~~.
\eeq
The assumption that the total orbital angular momentum of quarks in
a baryon vanishes entails the relation $A = A' = 1$ \cite{BR}.

The predictions of the na\"{\i}ve \cite{GR} and configuration-mixed
\cite{BR} models are compared with experimental values \cite{PDG}
in Table 4.  The configuration-mixed model was to be judged on the basis
of its prediction of a more negative $\Omega^-$ magnetic moment than the
na\"{\i}ve model's prediction $\mu(\Omega^-)
= 3 \mu (s) = 3 \mu (\Lambda)$.  The experimental
value is almost exactly between the two predictions -- if
anything, closer to the na\"{\i}ve number.

The configuration-mixed model of Ref.~\cite{BR} was probably an
oversimplification.  Deep inelastic scattering experiments tell us that
some of the proton's spin is carried by gluons or orbital angular
momentum and by sea quarks, so it is remarkable that the na\"{\i}ve model
works as well as it does.

The study of hyperon magnetic moments at Fermilab builds upon a
systematic investigation of hyperon polarizations, whose pattern is
still a puzzle for theorists.  We now discuss these results briefly.

\begin{table}
\caption{Baryon magnetic moments (in nuclear magnetons) in na\"{\i}ve
and configuration-mixed quark models, compared
with experimental values.}
\begin{center}
\begin{tabular}{c c c c}
Baryon & Na\"{\i}ve & Mixed & Experiment \\ \hline
$p$        &  Input  &  Input   &   2.793  \\
$n$        &  Input  &  Input   & $-1.913$ \\
$\Lambda$  &  Input  & $>-0.75$ & $-0.613 \pm 0.004$ \\
$\Sigma^+$ &   2.67  & $2.48 \pm 0.02$ & $2.458 \pm 0.010$ \\
$\Sigma^0 \to \Lambda$ & $-1.63$ & $>-1.78$ & $-1.61 \pm 0.08$ \\
$\Sigma^-$ & $-1.09$ &  Input   & $-1.160 \pm 0.025$ \\
$\Xi^0$    & $-1.44$ &  Input   & $-1.250 \pm 0.014$ \\
$\Xi^-$    & $-0.50$ &  Input   & $-0.6507 \pm 0.0025$ \\
$\Omega^-$ & $-1.84$ & $-2.26 \pm 0.09$ & $-2.02 \pm 0.05$ \\ \hline
\end{tabular}
\end{center}
\end{table}

\section{Hyperon Polarization}

A recent Tevatron fixed-target experiment on hyperon polarization,
from which earlier references may be traced, is Experiment 761
\cite{E761}.  We refer the reader to the original articles for
illustrations of the behavior of polarizations as functions of
Feynman $x_f$ and transverse momentum $P_t$, and merely describe the
pattern.  If the net spin of a hyperon is parallel to that
of the strange quark(s), as in the case of the $\Lambda$, $\Xi^0$,
and $\Xi^-$, the polarization is negative.  If, on the other hand,
the net spin is antiparallel to that of the strange quark, as in the
case of the $\Sigma^+$, the polarization is positive.  This behavior
was understood qualitatively twenty years ago in a
fragmentation model by DeGrand and Miettinen \cite{DM}.  However, the
polarization of {\it antihyperons} was not predicted in this model, and
the pattern so far has resisted explanation.  For example, the $\bar
\Sigma^-$ is produced with positive polarization, about half that of
its antiparticle, the $\Sigma^+$!  The $\bar \Xi^+$ is produced with
approximately the same polarization as the $\Xi^-$ \cite{Ho}!
Prediction of the pattern is another puzzle for future generations.
 
\section{The Unexpected}

The definition of ``unexpected'' depends on one's theoretical
predilections; it is the variety of these which leads to surprises.
I give just two examples.

\subsection{Neutral heavy leptons}

Right-handed neutrinos are natural in many schemes such as SO(10)
and its subgroup SO(6) $\otimes$ SO(4) which seek to unify the
electroweak and strong interactions.  Large Majorana masses $M$
of right-handed neutrinos don't violate any known symmetry, and
the lepton number violation which they entail is one candidate
\cite{BL} for the origin of the net baryon number of the Universe.

It appears that neutrinos have tiny masses, possibly smaller than
0.1 eV on the basis of atmospheric $\nu_\mu$ oscillations suggested
by experiments at SuperKamiokande \cite{atm}.  The seesaw model of
these masses \cite{SS} $m_\nu = m_{\rm Dirac}^2/M$ then implies that
right-handed neutrino masses $M$ must be above the reach of
conventional accelerator experiments, but it is wise to search anyway.
The NuTeV Collaboration \cite{RHN} has extended previous experimental
limits on masses and mixings of right-handed neutrinos produced, for
example, in decays of kaons and charmed particles, and has placed limits
\cite{NoK} on the production of a 33.6 MeV neutral lepton suggested by
another experiment \cite{KARMEN}.  Recently NuTeV has reported three
intriguing dimuon events from this search whose rate appears to exceed
background estimates \cite{Shaevitz}.  For one interpretation, see Ref.\
\cite{Borissov}.

\subsection{Unconventional families}

The repetitive family structure of the quarks and leptons is reminiscent
of the beginning of the periodic table of the elements.  Does it
suggest a composite structure for these objects?  Are new symmetries
involved?  As in the case of the periodic table, it may be necessary
to see variations in the pattern before its origin becomes clear.
Unification schemes based on groups beyond SO(10), such as $E_{\rm 6}$
\cite{E6}, predict such variations, entailing isosinglet quarks of
charge $-1/3$, vector-like (left-right symmetric) lepton multiplets,
and ``sterile'' (weak isosinglet)  neutrinos which need not have large
Majorana masses.

The LSND claim \cite{LSND} for $\bar \nu_\mu \to \bar \nu_e$
oscillations, when combined with evidence for solar and atmospheric
neutrino oscillations, probably requires at least one sterile
neutrino.  The large hierarchy between the $b$ and $t$ masses could
indicate that mixing between the $b$ and a heavier quark of charge
$-1/3$ is depressing the $b$ mass \cite{E6mix}.  We look forward
to future neutrino oscillation experiments at Fermilab \cite{MINOS,BooNE}
and elsewhere to elucidate the pattern of neutrino masses and
mixings, and to searches at high-energy colliders which may
uncover new states of matter.

\section{Conclusions}

The Fermilab Tevatron's fixed-target program has provided a superb 
variety and scope of experiments for nearly 20 years.  It has addressed
many issues through precision measurements, as is natural for
a facility working at the frontier of luminosity rather than the highest
center-of-mass energy.  Now we are entering an era of even more
precise and even lower-energy fixed target physics at Fermilab, to be
provided by the Main Injector.  We can look forward to exciting physics
from this program, in such areas as rare kaon decays, neutrino
oscillations, and -- we hope -- a generous dose of searches for the
unexpected.

\section*{Acknowledgements}

I wish to thank Joseph Lach and Mike Witherell for the invitation to
prepare this review, which was written in part at the Aspen Center for
Physics.  This work was supported in part by the United States
Department of Energy under Grant No.\ DE FG02 90ER40560.

\def \ajp#1#2#3{Am.\ J. Phys.\ {\bf#1}, #2 (#3)}
\def \apny#1#2#3{Ann.\ Phys.\ (N.Y.) {\bf#1}, #2 (#3)}
\def \app#1#2#3{Acta Phys.\ Polonica {\bf#1}, #2 (#3)}
\def \arnps#1#2#3{Ann.\ Rev.\ Nucl.\ Part.\ Sci.\ {\bf#1}, #2 (#3)}
\def \art{and references therein}
\def \cmts#1#2#3{Comments on Nucl.\ Part.\ Phys.\ {\bf#1}, #2 (#3)}
\def \cn{Collaboration}
\def \cp89{{\it CP Violation,} edited by C. Jarlskog (World Scientific,
Singapore, 1989)}
\def \efi{Enrico Fermi Institute Report No.\ }
\def \epjc#1#2#3{Eur.~Phys.~J.~C {\bf#1}, #2 (#3)}
\def \f79{{\it Proceedings of the 1979 International Symposium on Lepton and
Photon Interactions at High Energies,} Fermilab, August 23-29, 1979, ed. by
T. B. W. Kirk and H. D. I. Abarbanel (Fermi National Accelerator Laboratory,
Batavia, IL, 1979}
\def \hb87{{\it Proceeding of the 1987 International Symposium on Lepton and
Photon Interactions at High Energies,} Hamburg, 1987, ed. by W. Bartel
and R. R\"uckl (Nucl. Phys. B, Proc. Suppl., vol. 3) (North-Holland,
Amsterdam, 1988)}
\def \ib{{\it ibid.}~}
\def \ibj#1#2#3{~{\bf#1}, #2 (#3)}
\def \ichep72{{\it Proceedings of the XVI International Conference on High
Energy Physics}, Chicago and Batavia, Illinois, Sept. 6 -- 13, 1972,
edited by J. D. Jackson, A. Roberts, and R. Donaldson (Fermilab, Batavia,
IL, 1972)}
\def \ijmpa#1#2#3{Int. J. Mod. Phys. A {\bf#1}, #2 (#3)}
\def \ite{{\it et al.}}
\def \jhep#1#2#3{JHEP {\bf#1}, #2 (#3)}
\def \jpb#1#2#3{J.~Phys.~B~{\bf#1}, #2 (#3)}
\def \lg{{\it Proceedings of the XIXth International Symposium on
Lepton and Photon Interactions,} Stanford, California, August 9--14 1999,
edited by J. Jaros and M. Peskin (World Scientific, Singapore, 2000)}
\def \lkl87{{\it Selected Topics in Electroweak Interactions} (Proceedings of
the Second Lake Louise Institute on New Frontiers in Particle Physics, 15 --
21 February, 1987), edited by J. M. Cameron \ite~(World Scientific, Singapore,
1987)}
\def \kdvs#1#2#3{{Kong.~Danske Vid.~Selsk., Matt-fys.~Medd.} {\bf #1}, No.~#2
(#3)}
\def \ky85{{\it Proceedings of the International Symposium on Lepton and
Photon Interactions at High Energy,} Kyoto, Aug.~19-24, 1985, edited by M.
Konuma and K. Takahashi (Kyoto Univ., Kyoto, 1985)}
\def \mpla#1#2#3{Mod. Phys. Lett. A {\bf#1}, #2 (#3)}
\def \nat#1#2#3{Nature {\bf#1}, #2 (#3)}
\def \nc#1#2#3{Nuovo Cim. {\bf#1}, #2 (#3)}
\def \np#1#2#3{Nucl. Phys. {\bf#1}, #2 (#3)}
\def \npps#1#2#3{Nucl.\ Phys.\ Proc.\ Suppl.\ {\bf#1}, #2 (#3)}
\def \PDG{Particle Data Group, L. Montanet \ite, \prd{50}{1174}{1994}}
\def \pisma#1#2#3#4{Pis'ma Zh. Eksp. Teor. Fiz. {\bf#1}, #2 (#3) [JETP Lett.
{\bf#1}, #4 (#3)]}
\def \pl#1#2#3{Phys. Lett. {\bf#1}, #2 (#3)}
\def \pla#1#2#3{Phys. Lett. A {\bf#1}, #2 (#3)}
\def \plb#1#2#3{Phys. Lett. B {\bf#1}, #2 (#3)}
\def \pr#1#2#3{Phys. Rev. {\bf#1}, #2 (#3)}
\def \prc#1#2#3{Phys. Rev. C {\bf#1}, #2 (#3)}
\def \prd#1#2#3{Phys. Rev. D {\bf#1}, #2 (#3)}
\def \prl#1#2#3{Phys. Rev. Lett. {\bf#1}, #2 (#3)}
\def \prp#1#2#3{Phys. Rep. {\bf#1}, #2 (#3)}
\def \ptp#1#2#3{Prog. Theor. Phys. {\bf#1}, #2 (#3)}
\def \rmp#1#2#3{Rev. Mod. Phys. {\bf#1}, #2 (#3)}
\def \rp#1{~~~~~\ldots\ldots{\rm rp~}{#1}~~~~~}
\def \si90{25th International Conference on High Energy Physics, Singapore,
Aug. 2-8, 1990}
\def \slc87{{\it Proceedings of the Salt Lake City Meeting} (Division of
Particles and Fields, American Physical Society, Salt Lake City, Utah, 1987),
ed. by C. DeTar and J. S. Ball (World Scientific, Singapore, 1987)}
\def \slac89{{\it Proceedings of the XIVth International Symposium on
Lepton and Photon Interactions,} Stanford, California, 1989, edited by M.
Riordan (World Scientific, Singapore, 1990)}
\def \smass82{{\it Proceedings of the 1982 DPF Summer Study on Elementary
Particle Physics and Future Facilities}, Snowmass, Colorado, edited by R.
Donaldson, R. Gustafson, and F. Paige (World Scientific, Singapore, 1982)}
\def \smass90{{\it Research Directions for the Decade} (Proceedings of the
1990 Summer Study on High Energy Physics, June 25--July 13, Snowmass, Colorado),
edited by E. L. Berger (World Scientific, Singapore, 1992)}
\def \tasi{{\it Testing the Standard Model} (Proceedings of the 1990
Theoretical Advanced Study Institute in Elementary Particle Physics, Boulder,
Colorado, 3--27 June, 1990), edited by M. Cveti\v{c} and P. Langacker
(World Scientific, Singapore, 1991)}
\def \yaf#1#2#3#4{Yad. Fiz. {\bf#1}, #2 (#3) [Sov. J. Nucl. Phys. {\bf #1},
#4 (#3)]}
\def \zhetf#1#2#3#4#5#6{Zh. Eksp. Teor. Fiz. {\bf #1}, #2 (#3) [Sov. Phys. -
JETP {\bf #4}, #5 (#6)]}
\def \zpc#1#2#3{Zeit. Phys. C {\bf#1}, #2 (#3)}
\def \zpd#1#2#3{Zeit. Phys. D {\bf#1}, #2 (#3)}


\begin{thebibliography}{99}

\bibitem{Quigg} C. Quigg, ``The Electroweak Theory,'' in {\it Flavor Physics
for the Millennium (TASI 2000)}, edited by J. L. Rosner (World Scientific,
Singapore, 2001), p.\ 3.

\bibitem{DONUT} DONUT \cn~(Fermilab E-872), K. Kodama \ite,
\plb{504}{218}{2001}.

\bibitem{blife} MAC \cn, E. Fernandez \ite, \prl{51}{1022}{1983};
Mark II \cn, N. Lockyer \ite, \prl{51}{1316}{1983}.

\bibitem{WP} L. Wolfenstein, \prl{51}{1945}{1983}.

\bibitem{oldSig} P. Keller \ite, \prl{48}{971}{1982}, \art.

\bibitem{Cab} N. Cabibbo, \prl{10}{531}{1963}.

\bibitem{oldcharm} Fermilab E531 \cn, N. Ushida \ite, \prl{45}{1049,
1053}{1980}; \ibj{48}{844}{1982}; \ibj{51}{2362}{1983};
\ibj{56}{1767, 1771}{1986}.

\bibitem{GLR} M. K. Gaillard, B. W. Lee, and J. L. Rosner, \rmp{47}{277}
{1975}.

\bibitem{GLAM}
G. Altarelli, N. Cabibbo, and L. Maiani,
\np{B88}{285}{1975}; \pl{57B}{277}{1975}.

\bibitem{CCFT} J. H. Christenson, J. W. Cronin, V. L. Fitch, and R.
Turlay, \prl{13}{138}{1964}.

\bibitem{SW} L. Wolfenstein, \prl{13}{562}{1964}.

\bibitem{KM} M. Kobayashi and T. Maskawa, \ptp{49}{652}{1973}.

\bibitem{CKMrevs} For reviews of information on CKM matrix elements see,
e.g., F. Gilman, K. Kleinknecht, and Z. Renk, \epjc{15}{110}{2000}, in
{\it Review of Particle Properties}, D. E. Groom, \ite, \epjc{15}{1}{2000};
A. F. Falk, hep-ph/9908520, published in \lg, p.\ 174;
A. Ali and D. London, in {\it Proceedings of the Workshop on Physics and
Detectors for DAPHNE}, Frascati, Italy, Nov.\ 16--19, 1999, edited by S.
Bianco \ite~(INFN, Frascati, 1999), p.\ 3.

\bibitem{E715} Fermilab E715 \cn, S. Y. Hsueh \ite, \prl{54}{2399}
{1985}; \prd{38}{2056}{1988}.

\bibitem{Cumalat} J. Cumalat, ``The Status of Mixing in the Charm Sector,''
in {\it Flavor Physics for the Millennium} \cite{Quigg}, p.\ 243.

\bibitem{E832} KTeV (Fermilab E832) Collaboration, A. Alavi-Harati \ite,
\prl{83}{22}{1999}. A paper is in preparation reporting an updated value.
 
\bibitem{NA48} NA48 \cn, G. D. Barr \ite, presented at CERN seminar by A.
Ceccucci, Feb.\ 29, 2000 (unpublished).
An updated value was subsequently published:  A. Lai \ite, \epjc{22}{231}
{2001}.

\bibitem{EHLQ} E. Eichten, I. Hinchliffe, K. D. Lane, and C. Quigg,
\rmp{56}{579}{1984}; \ibj{58}{1065(E)}{1986}.

\bibitem{CCFR} See, e.g., J. M. Conrad, M. H. Shaevitz, and T. Bolton,
\rmp{70}{1341}{1998}, \art.

\bibitem{FT} {\it Symposium in Celebration of the Fixed Target Program
with the Tevatron}, Fermi National Accelerator Laboratory, June 2, 2000,
edited by J. A. Appel \ite~(Fermi National Accelerator Laboratory, 2000).

\bibitem{PW} E. A. Paschos and L. Wolfenstein, \prd{7}{91}{1973}.

\bibitem{NuTeV} NuTeV (Fermilab E815) \cn, G. P. Zeller \ite,
hep-ex/9906024, to be published in Proceedings of the American Physical
Society (APS) Meeting of the Division of Particles and Fields (DPF 99),
Los Angeles, CA, 5--9 January, 1999.  For an updated result see NuTeV \cn,
G. P. Zeller \ite, \prl{88}{091802}{2002}.

\bibitem{mt} G. Brooijmans (for Fermilab CDF and D0 \cn s),
FERMILAB-CONF-00-087-E, hep-ex/0005030; S. Leone (for Fermilab CDF and 
D0 \cn s) FERMILAB-CONF-00-115-E; CDF \cn, T. Affolder \ite,
\prd{63}{032003}{2001}.

\bibitem{bd} Fermilab Proposals E636 (I. A. Pless and T. Kitagaki,
spokesmen) and E646 (M. W. Peters, spokesman).

\bibitem{Kodama} Fermilab E653 \cn, K. Kodama \ite, \plb{382}{299} {1996}.

\bibitem{CLEODs} CLEO \cn, M. Chadha \ite, \prd{58}{032002}{1998}.

\bibitem{solex} See the experimental talks on solar neutrino
oscillations presented by A. McDonald, Y. Suzuki, V. Gavrin, E.
Bellotti, and K. Lande at the XIX International
Conference on Neutrino Physics and Astrophysics, Sudbury, Canada,
June 16--21, 2000.

\bibitem{solth} M. Gonzalez-Garcia, presented at XIX International
Conference on Neutrino Physics and Astrophysics, Sudbury, Canada,
June 16--21, 2000; M. Gonzalez-Garcia and C. Pena-Garay, \npps{91}{80}{2000}
[{\it Proceedings of Neutrino 2000}, edited by J. Law, R. W. Ollerhead, and J.
J. Simpson (North-Holland, Amsterdam, 2001)].

\bibitem{atm} H. Sobel, T. Mann, B. Barish, S. Mikheyev, P. Lipari,
and E. Lisi, talks presented at XIX International Conference on
Neutrino Physics and Astrophysics, Sudbury, Canada, June 16--21, 2000.

\bibitem{LSND} LSND \cn, presented by G. Mills
at XIX International Conference on Neutrino Physics and Astrophysics,
Sudbury, Canada, June 16--21, 2000.

\bibitem{MINOS} MINOS \cn, presented by S. Wojcicki
at XIX International Conference on Neutrino Physics and Astrophysics,
Sudbury, Canada, June 16--21, 2000.

\bibitem{BooNE} BooNE \cn, presented by A. Bazarko at
at XIX International Conference on Neutrino Physics and Astrophysics,
Sudbury, Canada, June 16--21, 2000.

\bibitem{GL} M. Gell-Mann and M. L\'evy, \nc{16}{705}{1960}.

\bibitem{Ups} S. W. Herb \ite, \prl{39}{252}{1977}; W. R. Innes
\ite, \prl{39}{1240, 1640(E)}{1977}.

\bibitem{top} CDF \cn, F. Abe \ite, \prd{50}{2966}{1994};
\prl{73}{225}{1994}; \ibj{74}{2626}{1995}; D0 \cn, S. Abachi \ite,
\prl{72}{2138}{1994}; \ibj{74}{2422, 2632}{1995}; \prd{52}{4877}{1995}.

\bibitem{BG} J. D. Bjorken and S. L. Glashow, \pl{11}{255}{1964}.

\bibitem{hypbeta} KTeV (Fermilab E799/E832) \cn, A. Affolder \ite,
\prl{82}{3751}{1999}.

\bibitem{JM} R. Flores-Mendieta, E. Jenkins, and A. V. Manohar,
\prd{58}{094028}{1998}.

\bibitem{SJ} J. L. Rosner, presented at 2nd Tropical Workshop on
Particle Physics and Cosmology, San Juan, Puerto Rico, May 1--6, 2000,
proceedings edited by J. F. Nieves (AIP, Melville, NY, 2000), p.\ 283.

\bibitem{E731} Fermilab E731 \cn, L. K. Gibbons \ite, \prl{70}{1203}
{1993}; \prd{55}{6625}{1997}.

\bibitem{NA31} CERN NA31 \cn, G. D. Barr \ite, \plb{317}{233}{1993}.

\bibitem{Buras} A. J. Buras, M. Jamin, and M. E. Lautenbacher, \plb{389}
{749}{1996}.

\bibitem{K99} See the articles on $\epsilon'/\epsilon$ by A. J. Buras,
S. Bertolini, R. Gupta, G. Martinelli, and W. A. Bardeen in {\it Kaon
Physics}, edited by J. L. Rosner and B. Winstein (University of Chicago
Press, 2000), p.\ 171.

\bibitem{BuK} G. Buchalla and A. J. Buras, \np{B548}{309}{1999};
G. Buchalla, in {\it Kaon Physics} \cite{K99}, p.\ 567.

\bibitem{E787} Brookhaven E787 \cn, S. Adler \ite, \prl{84}{3768}{2000}.
A further event was subsequently reported:  S. Adler \ite, \prl{88}{041803}
{2002}.

\bibitem{E949} See the description of Brookhaven National Laboratory
experiment E949 by L. Littenberg in {\it Kaon Physics} \cite{K99}, p.\ 575.

\bibitem{CKM} CKM \cn, Fermilab Proposal P905.

\bibitem{HG} H. B. Greenlee, \prd{42}{3724}{1990}.

\bibitem{pll} KTeV (Fermilab E-799) \cn, A. Alavi-Harati \ite,
presented by J. Whitmore at Kaon 99 Conference, Chicago, IL, June
21-26, 1999, published in {\it Kaon Physics} \cite{K99}, p.\ 415;
A. Alavi-Harati \ite, \prl{84}{5279}{2000}; \ibj{86}{397}{2001}.

\bibitem{Ko} See, however, the estimate by P. Ko, \prd{44}{139}{1991}.

\bibitem{K0pio} K0PIO \cn, in {\it Rare Symmetry Violating Processes},
proposal to the National Science Foundation, October 1999, and
Brookhaven National Laboratory Proposal P926 (unpublished).

\bibitem{KAMI} KAMI \cn, Fermilab Proposal P804 (unpublished).

\bibitem{BQ} KTeV (Fermilab E799) \cn, G. Breese Quinn, Ph.\ D. Thesis,
University of Chicago, May, 2000 (unpublished).

\bibitem{KTeVa} KTeV (Fermilab E799) \cn, A. Alavi-Harati \ite,
\prl{84}{408}{2000}.

\bibitem{NA48a} NA48 \cn, G. D. Barr \ite, presented by S. Wronka in
{\it Kaon Physics} \cite{K99}, p.\ 205.

\bibitem{SeS} L. M. Sehgal and M. Savage, articles to be published in
{\it Kaon Physics}, edited by J. L. Rosner and B. Winstein (University
of Chicago Press, 2000), pp.\ 181 and 189.

\bibitem{Pauli} See, e.g., the discussion by I. Bigi, M. Shifman, and
N. Uraltsev, \arnps{47}{591}{1997}, \art.

\bibitem{PDG} Particle Data Group, D. E. Groom \ite, \epjc{15}{1}{2000}.

\bibitem{CLEOmix} CLEO \cn, R. Godang \ite, \prl{84}{5038}{2000}.

\bibitem{FOCUS} FOCUS (Fermilab E831) \cn, J. M. Link \ite,
\plb{485}{62}{2000}.

\bibitem{ChTh} These are discussed, for example, by I. I. Bigi and N. G.
Uraltsev, \np{B592}{92}{2001};
S. Bergmann \ite, \plb{486}{418}{2000}.

\bibitem{GIM} S. L. Glashow, J. Iliopoulos, and L. Maiani, \prd{2}{1285}{1970}.

\bibitem{FOCUSCP} FOCUS (Fermilab E831) \cn, J. M. Link \ite,
\plb{491}{232}{2000}; \ibj{495}{443(E)}{2000}.

\bibitem{E605} Fermilab E605 \cn, G. Moreno \ite, \prd{43}{2815}{1991}.

\bibitem{Kwong} W. Kwong and J. L. Rosner, \prd{38}{279}{1988}.

\bibitem{Jpsiprod} CDF \cn, F. Abe \ite, \prl{69}{3704}{1992}; \ibj{79}{572}
{1997}; \ibj{79}{578}{1997}; D0 \cn, S. Abachi \ite, \plb{370}{239}{1996}.

\bibitem{EQ} S. D. Ellis, M. B. Einhorn, and C. Quigg, \prl{36}{1263}{1976}.

\bibitem{Braaten} P. Cho and M. B. Wise, \plb{346}{129}{1995};
E. Braaten and S. Fleming, \prl{74}{3327}{1995};
G. T. Bodwin, E. Braaten, and G. P. Lepage, \prd{51}
{1125}{1995}; \ibj{55}{5853(E)}{1997}; E. Braaten and Y.-Q. Chen,
\prd{54}{3216}{1996}; E. Braaten, S. Fleming, and
T. C. Yuan, \arnps{46}{196}{1996}.

\bibitem{Kniehl} B. A. Kniehl and J. Lee,
\prd{62}{114027}{2000}.

\bibitem{E771} Fermilab E771 \cn, S. Conetti and B. Cox, spokesmen.

\bibitem{E789} Fermilab E789 \cn, M. H. Schub \ite, \prd{52}{1307}
{1995}; \ibj{53}{570(E)}{1996}.

\bibitem{E789a} Fermilab E789 \cn, D. M. Jansen \ite, \prl{74}{3118}{1995}.

\bibitem{magmom} These include Fermilab Experiments No.\ 8, 361, 415,
440, 495, 619, 620, 756, and 800.

\bibitem{GR} See, e.g., S. Gasiorowicz and J. L. Rosner, \ajp{49}{954}
{1981}.

\bibitem{BS} L. Brekke and R. G. Sachs, \prd{28}{1178}{1983}; L. Brekke,
\apny{240}{400}{1995}.

\bibitem{BR} L. Brekke and J. L. Rosner, \cmts{18}{83}{1988}.

\bibitem{E761} Fermilab E761 \cn, A. Morelos \ite, \prl{71}{2172}{1995};
\prd{52}{3777}{1995}.

\bibitem{DM} T. A. DeGrand and H. I. Miettinen, \prd{23}{1227}{1981};
\ibj{24}{2419}{1981}; \ibj{31}{661(E)}{1985}; T. A. DeGrand,
J. Markkanen, and H. I. Miettinen, \prd{32}{2445}{1985}.

\bibitem{Ho} Fermilab E761 \cn, P. M. Ho \ite, \prl{65}{1713}{1990}.

\bibitem{BL} M. Fukugita and T. Yanagida, \plb{174}{45}{1986}; P. Langacker,
R. Peccei, and T. Yanagida, \mpla{1}{541}{1986}.

\bibitem{SS} M. Gell-Mann, P. Ramond, and R. Slansky, in {\it
Supergravity}, edited by P. van Nieuwenhuizen and D. Z. Freedman
North-Holland, Amsterdam, 1979), p.~315; T. Yanagida, in {\it
Proceedings of the Workshop on the Unified Theory and the Baryon
Number in the Universe}, edited by O. Sawada and S. Sugamoto,
KEK report KEK-79-18 (Tsukuba, 1979).

\bibitem{RHN} Fermilab E815, NuTeV \cn, A. Vaitaitis \ite, \prl{83}
{4943}{1999}.

\bibitem{NoK} NuTeV (Fermilab E815) \cn, J. A. Formaggio \ite,
\prl{84}{4043}{2000}.

\bibitem{KARMEN} KARMEN \cn, B. Armbruster \ite, \plb{348}{19}{1995}.

\bibitem{Shaevitz}  NuTeV (Fermilab E815) \cn, presented by M. Shaevitz at
XXXth International Conference on High Energy Physics, Osaka, Japan, July 27 --
August 2, 2000, Session PA 11e.

\bibitem{Borissov} L. Borissov, J. M. Conrad, and M. Shaevitz, Columbia
University Report, hep-ph/0007195 (unpublished).

\bibitem{E6} F. G\"ursey, P. Ramond, and P. Sikivie, \pl{60B}{177}{1976}.

\bibitem{E6mix} J. L. Rosner, \prd{61}{097303}{2000}.

\end{thebibliography}
\end{document}